\documentclass[journal]{IEEEtran}
\usepackage{epsfig,cite,mathptmx,times,amsmath,amssymb,color,flushend,gensymb,subfigure,
fancyhdr,subfigure,graphicx,cite,soul,framed,lipsum}

\usepackage{framed}
\definecolor{shadecolor}{rgb}{1,0.8,0.3}

\usepackage[switch,columnwise]{lineno}
\ifCLASSINFOpdf
\else
\fi
\begin{document}
\allowdisplaybreaks

\title{Experimental Validation of Linear and Nonlinear MPC on an Articulated Unmanned Ground Vehicle} 

\author{Erkan~Kayacan,~\IEEEmembership{Member, IEEE},~ Wouter~Saeys,~Herman~Ramon,~Calin~Belta,~\IEEEmembership{Fellow, IEEE}~and~Joshua~Peschel,~\IEEEmembership{Member, IEEE}

\thanks{This work was supported by the IWT-SBO 80032 (LeCoPro) project funded by the Institute for the Promotion of Innovation through Science and Technology in Flanders (IWT-Vlaanderen).}
\thanks{E. Kayacan is with Senseable City Laboratory and Computer Science $\&$ Artificial Intelligence Laboratory, Massachusetts Institute of Technology, Cambridge, Massachusetts 02139, USA. e-mail: {\tt\small erkank@mit.edu }}
\thanks{H. Ramon and W. Saeys are with the Division of Mechatronics, Biostatistics and Sensors, Department of Biosystems, University of Leuven (KU Leuven), 3001 Leuven, Belgium.
e-mail: {\tt\small \{herman.ramon, wouter.saeys\}@kuleuven.be }}
\thanks{C. Belta is with the Division of Systems Engineering, Boston University, Boston, MA, USA, cbelta@bu.edu.
e-mail: {\tt\small  cbelta@bu.edu }}
\thanks{J. Peschel is with the Department of Agricultural and Biosystems Engineering, Iowa State University, Ames, Iowa 50011-3270, USA. e-mail: {\tt\small peschel@iastate.edu }}
}

\markboth{\textbf{PREPRINT VERSION:} IEEE/ASME TRANSACTIONS ON MECHATRONICS, Volume 23, Issue 5, 2018.}
{Shell \MakeLowercase{\textit{et al.}}: Bare Demo of IEEEtran.cls for Journals}
\maketitle

\begin{abstract}

This paper focuses on the trajectory tracking control problem for an articulated unmanned ground vehicle. We propose and compare two approaches in terms of performance and computational complexity. The first uses a nonlinear mathematical model derived from first principles and combines a nonlinear model predictive controller (NMPC) with a nonlinear moving horizon estimator (NMHE) to produce a control strategy. The second is based on an input-state linearization (ISL) of the original model followed by linear model predictive control (LMPC). A fast real-time iteration scheme is proposed, implemented for the NMHE-NMPC framework and benchmarked against the ISL-LMPC framework, which is a traditional and cheap method. The experimental results for a time-based trajectory show that the NMHE-NMPC framework with the proposed real-time iteration scheme gives better trajectory tracking performance than the ISL-LMPC framework and the required computation time is feasible for real-time applications. Moreover, the ISL-LMPC produces results of a quality comparable to the NMHE-NMPC framework at a significantly reduced computational cost.

\end{abstract}


\IEEEpeerreviewmaketitle


\section{Introduction}

\IEEEPARstart{T}{he} size of arable farmland on the earth has been decreasing while human population has been increasing outstandingly. It is expected that the population of the world will reach $9.1$ billion by 2050. Therefore, agricultural production will have to be double in order to feed a larger population and provide increasing demands for bioenergy \cite{fao}. To meet the demand for agricultural products, one possible solution is the automation of agricultural machines to get higher efficiencies and better precisions. Moreover, multitasking operations are needed in agricultural applications. For instance, a human operator simultaneously has to drive the agricultural vehicle with high precision, and adjust the position of a trailer and/or further parameters of several agricultural apparatus during tillage and planting. In this instance, a sophisticated and versatile control algorithm for the navigation of unmanned agricultural ground vehicles is a necessity to lead to an additional increment in the performance of the human operator.

The current implementations for automatic guidance of autonomous ground vehicles are based on either local positioning systems (vision or laser-based sensors) or global positioning systems (GPSs). Local positioning systems have been used in autonomous applications since the 70s \cite{julian1971design,reid1987vision}. It has been reported that their main disadvantage is the sensitivity to light conditions in outdoor environments although they are cheap to implement \cite{hiremath2014laser}. Recent developments in satellite technologies have led to an increase in the use of the latter, which has gradually replaced the former prevalent in the 90s \cite{bradford1996global,bell2000automatic}. Real-time kinematic (RTK) GPS yielding centimeter precision \cite{Bevly2007} has enabled intensive research on agricultural vehicles. Automated agricultural vehicles with GPSs have many advantages, such as extricating the driver from tiresome tasks of accurately steering the vehicle, increasing trajectory tracking accuracy, and being able to operate at night or in foggy weather.

Various control techniques have been used to solve the trajectory tracking problem for tractors with and without trailers \cite{Khalaji2014,Michalek2015}. An adaptive controller was employed for a tractor assembled with dissimilar trailers in order to track straight lines \cite{Derrick}. Moreover, a linear optimal control method was proposed for a tractor-trailer system \cite{karkeejournal}. These controllers have been contingent on linearized dynamic and kinematic models, which are only valid for small yaw deviations around a fixed value and small steering angles, such that they are restricted to linear trajectories.

Model predictive control (MPC) is a popular technique in the process industry for multi-input-multi-output applications \cite{7505627, Kayacan2018}. Forasmuch as the tractor-trailer system can be described by variable set points for following curvilinear trajectories, this involves a merger between MPC structure and a nonlinear model as known nonlinear MPC (NMPC). NMPC was designed for a tractor-trailer system along with a curvilinear trajectory in \cite{Backman2012} while an extended Kalman filter (EKF) was designed to estimate of the yaw angles of the tractor and trailer. In \cite{TomKraus}, the states of an agricultural vehicle including slips parameters were estimated with nonlinear moving horizon estimation (NMHE) and forwarded to an NMPC. This concept has been extended for the tracking of a space-based trajectory by a tractor-trailer system in a centralized control structure and accomplished results have been reported in \cite{erkanCeNMPC}. Moreover, decentralized and distributed NMPC approaches have been recommended to reduce the computational burden with minimal loss of tracking performance \cite{erkanDeNMPC,erkanDiNMPC} .

Although trajectory tracking performance obtained in the aforementioned studies is quite good, traditional NMPC implementations are computationally expensive. Therefore, the aim of this study was to design fast frameworks for trajectory tracking problem and compare their performance with regard to computational burden and tracking error. Firstly, a fast NMHE-NMPC framework is designed for tracking a time-based trajectory. In the NMHE-NMPC framework, the NMHE learns traction parameters using onboard sensors online, and the NMPC enables high accurate trajectory tracking. Thus, we provide robust tracking performance when uncertainty is high as uncertainty is reduced through learning whereas traditional NMPC approaches do not typically account for model uncertainty. Moreover, a real-time iteration scheme is proposed to solve NMHE and NMPC problems efficiently. Secondly, it is shown that the nonlinear model is input-state linearizable and an LMPC is proposed for the linear transformation of the system. Both the frameworks are then implemented on a real-time system and benchmarked against each other.

This paper is organized as follows: The system is described in Section \ref{sectionsystem}. The formulations and implementations of NMHE and NMPC are given in Section \ref{nmhenmpcframework}. The input-state linearization approach and LMPC control structure are explained in Section \ref{isllmpcframework}. The real-time experimental results are given in Section \ref{experimentalresults}. Finally, the study is summarized in Section \ref{conclusions}.


\section{Unmanned Tractor-Trailer System}\label{sectionsystem}

\begin{figure}[b!]
\centering
  \includegraphics[width=3.4in]{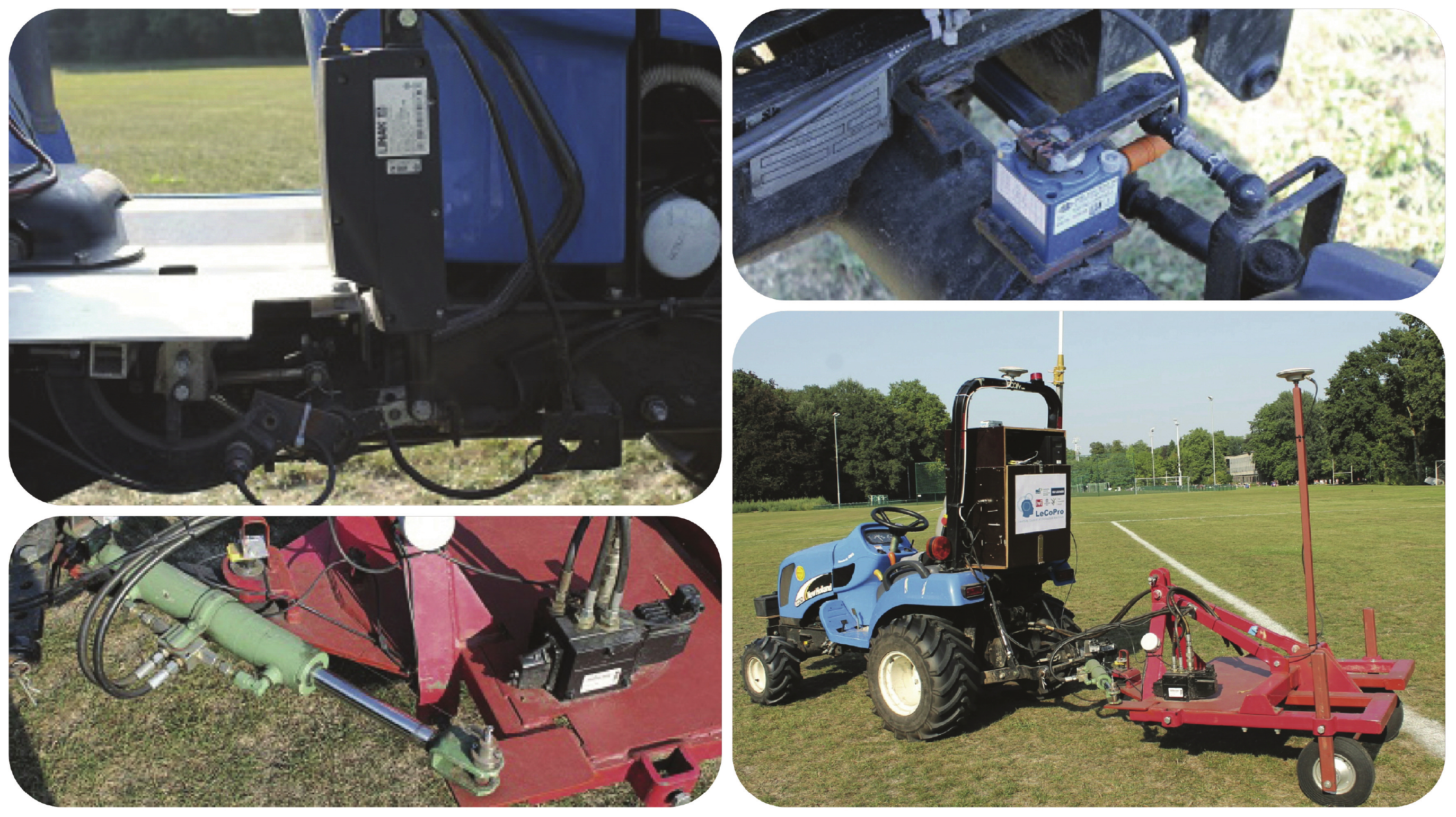}\\
  \caption{On the top left side: the electro-mechanical valve, on the bottom left side: the electro-hydraulic valve, on the top right side: the potentiometer,  on the bottom right side: the articulated unmanned ground vehicle.}
  \label{tractor-trailer}
\end{figure}

The goal of this paper is to obtain a precise trajectory tracking performance to ensure constant distances between rows to prevent from crop damage while variable soil conditions are subjected to an uneven, rough and wet grass field. The small tractor-trailer system, the actuators, and the sensors are shown in Fig. \ref{tractor-trailer}. 

In order to measure the global position of the system, a real-time kinematic global positioning system (AsteRx2eH, Septentrio Satellite Navigation NV, Belgium) with the aid of the Flepos network is used with two antennas installed on the tractor and the trailer. There are three actuators to control the system: two electro-hydraulic valves (OSPC50-LS/EH-20, Dan-foss, Nordborg, Denmark) for the steering mechanisms and an electromechanical valve (LA12, Linak, Nordborg, Denmark) for the hydrostat system. In addition, a potentiometer (533-540- J00A3X0-0, Mobil Elektronik, Langenbeutingen, Germany), an inductive sensor and two encoders mounted on the rear wheels are used to measure respectively the angle of the front wheels of the tractor, the steering angle of the trailer, and the speed of the system. A real time operating system equipped with a 2.26 GHz Intel $Core^{TM}$2 Quad Q9100 quad-core processor is used to implement the control algorithms that have been executed and updated at a rate of 200 milliseconds in $LabVIEW^{TM}$.

\begin{figure}[t!]
\centering
 \includegraphics[width=\columnwidth]{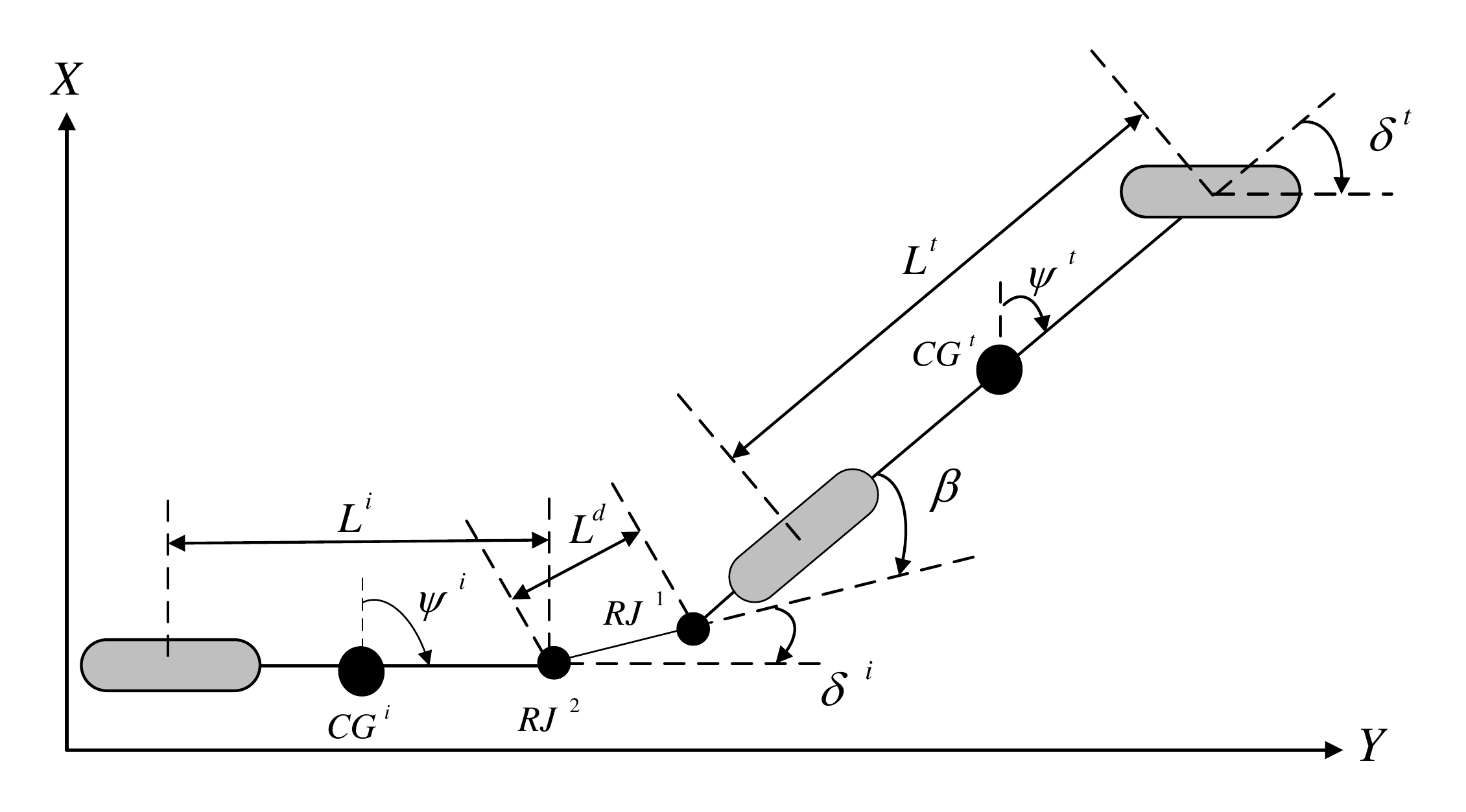}\\
  \caption{A schematic representation of tricycle model for an unmanned tractor-trailer system}
  \label{kinematic}
\end{figure}
The adaptive kinematic model for the unmanned tractor-trailer system is an extended version of the one presented in \cite{karkeejournal, karkeephd}. In order to make the system model adaptive, three traction parameters ($\mu$, $\kappa$, $\eta$) are inserted. The tractor and trailer rigid bodies are mechanically coupled by the drawbar so that there exist two revolute joints (RJs) that interconnect the drawbar to the tractor at $RJ^{1}$ and the drawbar to the trailer at $RJ^2$ as illustrated in Fig. \ref{kinematic} . The centers of gravity of the tractor and trailer are respectively represented by $CG^t$ and $CG^i$.
The equations for the system, which are a combination of the kinematic model in \cite{erkan2016mech} and the speed model in \cite{erkanmodelleme}, are written:
\begin{eqnarray}\label{kinematicmodel}
  \dot{x}^{t} & = &  \mu v \cos{(\psi^{t})} \nonumber \\
  \dot{y}^{t} & = & \mu v \sin{(\psi^{t})} \nonumber \\
  \dot{\psi}^{t} & = &\frac{\mu v \tan{ (\kappa \delta^{t}) } }{L^t} \nonumber \\
  \dot{x}^{i}  & = & \mu v \cos{(\psi^{i})} \nonumber \\
  \dot{y}^{i} & = & \mu v \sin{(\psi^{i})} \nonumber \\
  \dot{\psi}^{i}  & = & \frac{\mu v }{L^i}\Big(\sin{(\eta \delta^{i} + \beta)} + \frac{L^{d} }{L^t} \tan{ (\kappa \delta^{t})} \cos{(\eta \delta^{i} + \beta)} \Big) \nonumber \\
  \dot{v} & = &-\frac{v}{\tau} + \frac{K}{\tau} HP
\end{eqnarray}
where ${x}^{t}$ $(m)$, $y^{t}$  $(m)$, ${x}^{i}$  $(m)$, $y^{i}$ $(m)$, $\psi^{t}$  $(rad)$ and $\psi^{i}$ $(rad)$ denote respectively the positions and yaw angles of the tractor and trailer while $v$ $(m/s)$ denotes circumferential speed of the wheels. The steering angles are respectively denoted by $\delta^{t}$ $(rad)$ and $\delta^{i}$ $(rad)$ for the tractor and trailer while the angle at $RJ^{1}$ and the hydrostat position are respectively denoted by $\beta$ $(rad)$ and $HP$ $(\%)$. Moreover, $\mu$, $\kappa$ and $\eta$ denote for the traction coefficients for the longitudinal and side slips. It should be noted that the traction parameters can merely acquire values between zero and one on asphalt roads and soil surfaces. If the traction parameter for the longitudinal slip is equal to one, all the rotary motion of the wheels is transformed into the linear motion of the vehicle. Moreover, a proportion of zero expresses that tires are rotating and the ground wheel speed is equal to zero so that the system is not completely controllable. In other words, the definition of the traction parameter $\mu$ for longitudinal slip allows to convert the circumferential wheel speed $\nu$ into a \emph{ground wheel speed $\mu \nu$}. The definition of the two traction parameters for side-slips maintains a manner to determine \emph{effective steering angles}, i.e., $\kappa \delta^{t}$ and $\eta \delta^{i}$. It is assumed that each fraction for each steering angle ensures the actual tractor and trailer turning motions. Furthermore, the difference between yaw angles is equal to the summation of the angle at $RJ^1$ and the steering angle of the trailer, i.e.,  $\phi^i - \phi^t = \beta + \delta^i$. Since there are constraints on $\beta$ and $\delta^i$ defined respectively in \eqref{nmhe} and \eqref{nmpc}, it is to be noted that the yaw angle difference cannot become larger than $ \beta + \delta^i = 45$ degrees.

The equations in \eqref{kinematicmodel} are formulated in the following form;
\begin{eqnarray}\label{f_function}
\dot{x}  =  f \big(x,u,p \big) \quad and \quad
y  =  h \big(x,u,p \big)
\end{eqnarray}
with
\begin{eqnarray}\label{xstate}
x & = & \left[
  \begin{array}{ccccccc}
   x^{t} & y^{t} & \psi^{t} & x^{i} & y^{i} & \psi^{i} & v
  \end{array}
  \right]^{T} \\ \label{uinput}
u & = & \left[
  \begin{array}{ccc}
   \delta^t & \delta^i & HP
  \end{array}
  \right]^T \\ \label{parameter}
  p & = & \left[
  \begin{array}{cccc}
   \mu & \kappa & \eta & \beta
  \end{array}
  \right]^T \\ \label{youtput}
y & = & \left[
  \begin{array}{ccccccccc}
   x^{t} & y^{t} & x^{i} & y^{i} & v & \delta^t & \delta^i  & HP &  \beta
  \end{array}
  \right]^{T}
\end{eqnarray}
where x, u, p and y denote respectively the vectors of state, input, varying parameter and output of the system. 
The measured, fixed physical parameters are the distance between the front and rear wheels of the tractor $L^{t}=1.4m$, the distance between the $RJ^{2}$ and the rear wheel of the trailer $L^{i}=1.3m$ and the distance between the rear wheel of the tractor and $RJ^{2}$ $L^{d}=1.1m$. The identified, fixed parameters are \cite{erkanmodelleme}:  the time-constant $\tau= 2.05$ and the gain $K = 0.016$ for the wheel speed model while the engine speed is at $2500$ RPM. The angle between the tractor and drawbar $\beta$ is measured and the traction parameters $\mu, \kappa, \eta$ are estimated online so that the parameters in \eqref{parameter} can vary over time.


\section{NMHE-NMPC Framework}\label{nmhenmpcframework} 

NMHE-NMPC framework was develoepd and implemented for the space-based trajectory approach in \cite{Kayacan2018,erkanCeNMPC, erkanDeNMPC}. In this section, we will develop this framework for the time-based trajectory approach. 

\subsection{Nonlinear Moving Horizon Estimation}

Although values of all system states must be gathered for NMPC, it is impossible to measure all of them in practice. For this reason, it is a requirement to estimate unmeasured states or unknown model parameters online. The traditional method as a state estimator is the Extended Kalman filter (EKF). However, the major drawback of the EKF is that it cannot take the bounds on the states into account. To accomplish this restraint of the EKF in this paper, NMHE has been employed inasmuch as it takes the state and parameter estimation regarding bounds into account within the same problem \cite{TomKraus,Rao2003,Kuhl2011}.

In this paper, we consider an NMHE formulation in the following form at each sampling time t:
\begin{equation}
 \begin{aligned}
 & \underset{x(.),p,u(.)}{\text{min}}
 & & \left\|
  \begin{array}{c}
    \hat{x} (t_k-t_h) - x (t_k-t_h)  \\
    \hat{p} - p
 \end{array}
 \right\| ^{2}_{P}   + \int^{t_k}_{t_k-t_h} \| y_m - y (t) \|^{2}_{H} dt  \\
 & \text{s. t.}
 &&  \dot{x}(t) = f \big(x(t),u(t),p \big) \\
 &&&  y(t) = h \big(x(t),u(t),p \big) \\
  &&& -20 \deg  \leq \beta \leq 20 \deg  \\
 &&& 0 \leq \mu, \kappa, \eta \leq 1 \quad \forall t \in [t_k - t_h ,t_k]
  \end{aligned}
    \label{nmhe}
\end{equation}
where $y_m$ and $y$ denote respectively the measured output and the output function of the system model. The deviations in the estimates for the states and parameters before the estimation horizon $\hat{x} (t_k-t_h)$ and $\hat{p} (t_k-t_h)$ are minimized by a symmetric positive semi-definite weighting matrix $P$, while the deviations of the measured and system outputs in the estimation horizon are minimized by a symmetric positive semi-definite weighting matrix $H$ \cite{Ferreau}. The first part of the cost function in \eqref{nmhe} is named the arrival cost and must be bounded as a requirement. If not, it may go to infinity. Therefore, the impact of the old measurements on P is reduced by a weighting matrix $D_{update}$ in \eqref{Dupdate} \cite{Kuhl2011}.

The NMHE method can estimate the immeasurable states and parameters of the system  model simultaneously. The parameters have been assumed to be time-invariant and not subject to process noise over the estimation horizon. However, it is assumed that the parameters are time-varying Gaussian random variables in the arrival cost. Therefore, additional weighting factors must be added as the variance of the parameters noise and  the parameters appears only in the arrival cost. The extended weighting matrix $D_{update} \in R^{(n_x+n_p) \times(n_x+n_p)}$ can be written as follows:
\begin{eqnarray}\label{Dupdate}
D_{update}= \left[ \begin{array}{cc}
D^{n_{x}} & 0 \\
0 & D^{n_{p}} \\
\end{array}
\right]
\end{eqnarray}
where $D^{n_{x}} \in R^{n_{x} \times n_{x}}$ and $ D^{n_{p}} \in R^{n_{p} \times n_{p}}$ represent the weighting matrix for the state noise covariance and the weighting matrix for the parameter pseudo-variance. The weighting matrix $D_{update}$ is chosen based on the objective. Low gain in the process noise results in better estimation accuracy; however, it causes time-lag between true and estimated values. Therefore, the weighting coefficients for the measured states and parameters (e.g., $x^{t}$, $y^{t}$, $x^{i}$, $y^{i}$, $v$ and $\beta$) are selected large while the weighting coefficients for the immeasurable states and parameters (e.g., $\psi^{t}$, $\psi^{i}$, $\mu$, $\kappa$ and $\eta$) are selected small in this paper. Thus, the input to  the NMHE  algorithm becomes the output of  the system in \eqref{youtput} while the output of NMHE becomes the full states in \eqref{xstate} and full varying parameters in \eqref{parameter}. Moreover, the standard deviations of the measurements have been set to $\sigma_{x^{t}}=\sigma_{y^{t}}=\sigma_{x^{i}}=\sigma_{y^{i}}=0.03$ m, $\sigma_{\beta}= 0.0175$ rad, $\sigma_v = 0.1$ m/s, $\sigma_{\delta^{t}} = 0.0175$ rad, $\sigma_{\delta^{i}} = 0.0175$ rad and $\sigma_{HP} = 3$ based on the information obtained from the real- time experiments. The following weighting matrices $H$ and $D_{update}$  have been used in NMHE:
\begin{eqnarray}\label{weightingmatricesHD}
H &=& diag(\sigma^{2}_{x^t},\sigma^{2}_{y^t},\sigma^{2}_{x^i},\sigma^{2}_{y^i},
\sigma^{2}_{v}, \sigma^{2}_{\delta^{t}}, \sigma^{2}_{\delta^{i}}, \sigma^{2}_{HP}, \sigma^{2}_{\beta})^{-1} \nonumber  \\
D_{update} &=& diag(x^t, y^t, \psi^{t}, x^i, y^i, \psi^{i}, \mu, \kappa, \eta, \beta, v) \nonumber \\
      &=& diag(10.0, 10.0, 0.1, 10.0, 10.0, 0.1, \nonumber \\
      && 0.25, 0.25, 0.25, 1, 1)
\end{eqnarray}
The estimation horizon $t_{h}$ has been set to 3 seconds.

\subsection{Nonlinear Model Predictive Control}\label{sectionnmpc}

A nonlinear model represented in \eqref{f_function} $f(\cdot,\cdot,\cdot): \mathbb{R}^{n_{x}} \times \mathbb{R}^{n_{u}} \longrightarrow \mathbb{R}^{n_{x}} $ is the continuously state update function and $f(0,0,p)=0 \quad \forall t$ in which $x$ $\in$ $\mathbb{R}^{n_{x}}$ and $u$ $\in$ $\mathbb{R}^{n_{u}}$ are the state and input vectors. The states and inputs have to fulfill $ x \in \mathbb{X}, \;\;\; u \in \mathbb{U} $ where $\mathbb{X} \subseteq  \mathbb{R}^{n_{x}}$ is closed, $\mathbb{U} \subseteq  \mathbb{R}^{n_{u}}$ is compact and each set contains the origin in its interior point. 

In this study, we consider an NMPC formulation at each sampling time $t$ in the following form:
\begin{equation}
 \begin{aligned}
 & \underset{x(.), u(.)}{\text{min}}
 & & \int^{t_k+t_h}_{t_k} \Big( \| x_{r} (t) - x (t) \|^{2}_{Q} + \| \Delta u (t)  \|^{2}_{R} \Big) dt  \\
 &&& + \| x_{r} (t_k+t_h) - x (t_k+t_h) \|^{2}_{S} \\ 
 & \text{s.t.}
 && x(t_k) = \hat{x} (t_k) \\
 && & \dot{x}(t) = f \big(x(t), u(t), p \big) \\
 && & -35 \deg  \leq  \delta^{t}(t)  \leq 35 \deg \ \\
 && &-25 \deg  \leq \delta^{i}(t)  \leq  25 \deg  \\
 && & 0 \%  \leq  HP (t)  \leq  100 \%  \qquad \forall t \in [t_k, t_k+t_h]  
  \end{aligned}
  \label{nmpc}
\end{equation}
where the first and last parts are called the stage cost and the terminal penalty enforced the stability of NMPC in \cite{Mayne} in which $Q \in \mathbb{R}^{n_{x} \times
 n_{x}}$, $R \in \mathbb{R}^{n_{u} \times n_{u}}$ and $S \in \mathbb{R}^{n_{x} \times n_{x}}$ are symmetric positive definite weighting matrices, $x_{r}$ denotes respectively the references for the states, $x$ and $\Delta u$ denote respectively the states and the change of the inputs, $t_k$ denotes the current time, $t_h$ denotes the prediction horizon. $\hat{x} (t_k)$ denotes the estimated state vector by the NMHE. The first sample of $u(t)$, $u(t,x(t))= u^*(t_{k})$, is applied to the system and the NMPC problem is solved again over a moving horizon for the subsequent sampling time \cite{Kayacan2018}.

The references for the state are written:
\begin{eqnarray}\label{}
x_{r} & = & (x^t_{r},y^t_{r}, \psi^{t}_{r}, x^i_{r},y^i_{r}, \psi^{i}_{r}, v_{r})^T 
\end{eqnarray}

The weighting matrices Q, R and S have been written:
\begin{eqnarray}\label{weightingmatricesQRS}
Q & = & diag(2,2,0,4,4,0, 0), \quad S = 10 \times Q \nonumber \\
R & = & diag(7,7,7)
\end{eqnarray}

The weighting matrix R is selected larger than the weighting matrix Q so as to obtain well damped closed-loop system response. The other justification is that the system dynamics are slow so that it is not able to give a rapid reaction. Inasmuch as the last state error value in the prediction horizon is so crucial for the stability issues, the weighting matrix S is adjusted to 10 times larger than the weighting matrix Q. 

If the prediction and control horizons are selected large, the computation burden for NMPC will increase unreasonably so that solving the optimization problem will be infeasible. Moreover, if the prediction and control horizons are selected too small, the stabilization of the system may not be achieved. As reported in \cite{Falcone2007}, the prediction and control horizons of the NMPC must be large enough for a stable performance taking the velocity of the vehicle into consideration. Since the velocity of the tractor-trailer system is quite low, the prediction and control horizons do not have to be very large in this study. Therefore, the prediction and control horizons $t_h$ have been set to 3 seconds.

\subsection{Implementation}

The optimization problems in NMHE \eqref{nmhe} and NMPC \eqref{nmpc} are very similar so that using the same solution method for both of them makes sense \cite{TomKraus}. Inasmuch as they are nonlinear and non-convex optimization problems, the computational burden for solving these problems is quite large, and depends on the order of the system, the non-linearity of the system, the horizon length and used nonlinear optimization solver. 

In this study, the multiple shooting method has been consolidated with a generalized Gauss-Newton method \cite{Diehl2005}. The significant benefit is that the second derivatives that are arduous computing are not necessary. However, the drawback is that it is troublesome to foreknow the required number of iterations to attain a desired accuracy \cite{Houska2011a}. A simple solution that limits the number of iterations to 1 was proposed in \cite{Diehl}. Moreover, the Gauss-Newton iteration is divided into two parts: preparation and feedback parts. The preparation part is executed prior to the feedback part, and the feedback part is executed after measurements for NMHE and estimates for NMPC are available. In the preparation part, the system dynamics are integrated with the previous solution, and objectives, constraints, and corresponding sensitives are evaluated. In the feedback part, a single quadratic programming is solved with the current measurements for the NMHE and the current estimates for the NMPC. Thus, the new estimates for the NMHE and a new control signal for the NMPC are obtained. Compared to the classical method, this method minimizes feedback delay and produces similar results with higher computational efficiency \cite{Diehl}. Furthermore, the NMPC and NMHE are run in parallel, on separate processor cores, the NMHE preparation step is triggered at the same time as the NMPC feedback step. Therefore, this solution method reduces the overall required time for the preparation steps of the NMHE-NMPC. The ACADO code generation tool has been used to solve the constrained nonlinear optimization problems in the NMPC and NMHE \cite{Houska2011a}. Moreover, qpOASES software package, which is an open-source C++ implementation of online active set strategy, has been used as a QP solver \cite{Ferreau2014}. 


\section{ISL-LMPC Framework}\label{ISL-LMPC}\label{isllmpcframework}
\subsection{Input-State Linearization Transformation}

In the ISL-LMPC framework, an EKF has been used as an estimator. Since an EKF is not able to deal with the bounds on the states and parameters, we exclude the traction parameters for this framework. 

The input-state linearization is a useful method to compensate the non-linearity of a system. In this section, a nonlinear model for the tractor-trailer system excluding the traction parameters in \eqref{kinematicmodel} is transformed into a virtual linear model by using an input-state linearization method. Once a virtual linear model has been obtained, linear control techniques are used to design a controller for the overall system.

By taking a nonlinear system in \eqref{f_function} into account in which $f(x(t),u(t),p)$ is input-state linearizable if there exists a diffeomorphism such that the new state variables $z=T_x(x)$ transform the nonlinear system in \eqref{f_function} into the following linear time-invariant system \cite{slotine}:
\begin{equation}
\dot{z}=Az+B u_{z}
  \label{zlinearmodel}
\end{equation}
where the pair (A,B) is controllable. The transformation between the real and virtual control inputs resulting in the compensation of the system nonlinearities and a controllable linear system can be written as follows:
\begin{equation}
u= \phi(x) + T_{u}(x) u_{z}
  \label{virtualcontrolinput}
\end{equation}
where $T_{u}(x)$ is assumed to be non-singular \cite{slotine}.

The new states, the positions and velocities of the tractor and trailer, are defined as follows:
\small
\begin{eqnarray}\label{zstate}
z & = & \left[
  \begin{array}{cccccccc}
      x^{t} & y^{t} & v \cos{\psi^{t}} & v \sin{\psi^{t}}  &  x^{i} & y^{i} & v \cos{\psi^{i}} & v \sin{\psi^{i}}
  \end{array}
  \right]^{T} \nonumber \\
  & = & \left[
  \begin{array}{cccccccc}
   z_{1} & z_{2} & z_{3} & z_{4} & z_{5} & z_{6} & z_{7} & z_{8}
  \end{array}
  \right]^{T}
\end{eqnarray}
\normalsize

By combing the time derivative of \eqref{zstate} with the equations for the yaw angles and longitudinal speed model in \eqref{kinematicmodel}, the state-space model can be written as follows:
\begin{eqnarray}\label{zstatemodel}
\dot{z} & = & A z + B u_{z} \\ \label{zoutputmodel}
y_{z} & = & C z
\end{eqnarray}
where $\dot{z}_{1}=\dot{z}_{3}, \dot{z}_{2}=\dot{z}_{4}, \dot{z}_{3}=-z_{3}/\tau + u_{z_{1}},  \dot{z}_{4}=-z_{4}/\tau + u_{z_{2}}, 
\dot{z}_{5}=\dot{z}_{7}, \dot{z}_{6}=\dot{z}_{8}, \dot{z}_{7}= -z_{7}/\tau + u_{z_{3}}, \dot{z}_{8}=-z_{8}/\tau + u_{z_{4}} $ and $y_{z}= \left[
  \begin{array}{cccc}
  z_{1} & z_{2}  & z_{5} & z_{6}
  \end{array}  \right]^T$

As can be seen from the formulation above, there are 4 inputs for the virtual linear system even though the number of inputs for the real-time system is equal to 3. This results in 2 input transformations for the hydrostat position HP. One of these transformations is based on the position of the tractor while the other is calculated with respect to the information coming from the trailer. Since the hydrostat position HP is the input for the speed measured by encoders mounted on the tractor rear wheels, the transformation obtained from the equations of the tractor is used for the hydrostat position transformation. Moreover, the steering angle of the trailer is not input-linearizable for the transformation. Therefore, we have to rely on the small steering angle assumption so that the term $\cos(\delta^{i} + \beta)$ is assumed to be equal 1. Thus, the total input transformation can be written as follows:
\begin{eqnarray}\label{}
\delta^{t} & = & \arctan{\Big( \frac{L^{t} (-u_{z_{1}} \sin{\psi^{t}} + u_{z_{2}} \cos{\psi^{t}} )}{ v^{2}} \Big)} \nonumber \\
\delta^{i} & = & \arcsin \Big( \frac{L^{i} (-u_{z_{3}} \sin{\psi^{i}} + u_{z_{4}} \cos{\psi^{i}} )}{{ v^{2}} } \nonumber \\
 && - \frac{L^{d}}{L^{t}} (-u_{z_{1}} \sin{\psi^{t}} + u_{z_{2}} \cos{\psi^{t}} )  \Big)   - \beta \nonumber \\
HP & = & \frac{\tau}{K}\Big( u_{z_{1}} \cos{\psi^{t}} + u_{z_{2}} \sin{\psi^{t}} + \frac{v}{\tau} \Big)
\end{eqnarray}

\subsection{Linear Model Predictive Control}\label{sectionlmpc}

In this study, we considered the following LMPC formulation at each sampling time $t$:
\begin{equation}
 \begin{aligned}
 & \underset{x(.), u(.)}{\text{min}}
 & & \int^{t_k+t_h}_{t_k} (\| z_r (t) - z (t) \|^{2}_{Q}  +  \| \triangle u_{z} (t) \|^{2}_{R}) dt  \\
 &&&  +  \| z_r (t_k+t_h)  - z  (t_k+t_h)  \|^{2}_{S}  \\
 & \text{s. t.}
 && \dot{z}(t) = A z(t) + B u_{z}(t) \\
  && & -2 \leq z_{3} (t), z_{4} (t), z_{7}(t),  z_{8}(t) \leq 2 \quad \forall t \in [t_k, t_k+t_h] \\
  && & -2 \leq z_{3} (t_{k} + t_{h}), z_{4} (t_{k} + t_{h}), z_{7}(t_{k} + t_{h}),  z_{8}(t_{k} + t_{h}) \leq 2 
  \end{aligned}
  \label{lmpc}
\end{equation}
where $z_r$ is the references for the system states and $\triangle u_{z}$ is the change of the input. The maximum speed of the system is $2$ $m/s$; therefore, the constraints on $z_{3}$, $z_{4}$, $z_{7}$ and $z_{8}$ are defined in the formulation of the LMPC.

The prediction horizon $t_h$ has been set to $3$ second. As motivated in Section \ref{sectionnmpc}, the prediction horizon must not be very large due to the fact that the velocity of the system is too low. Moreover, the weighting matrices $Q$, $R$ and $S$ have been defined as follows:
\begin{eqnarray}\label{}
Q  &=&  diag(1,1,0,0,0.01,0.0,0,0),  \quad S  = 10 \times Q  \nonumber \\
R  &=&  diag(1,1,0.01,0.01) 
\end{eqnarray}

\begin{figure}[t!]
\centering
  \includegraphics[width=3.6in]{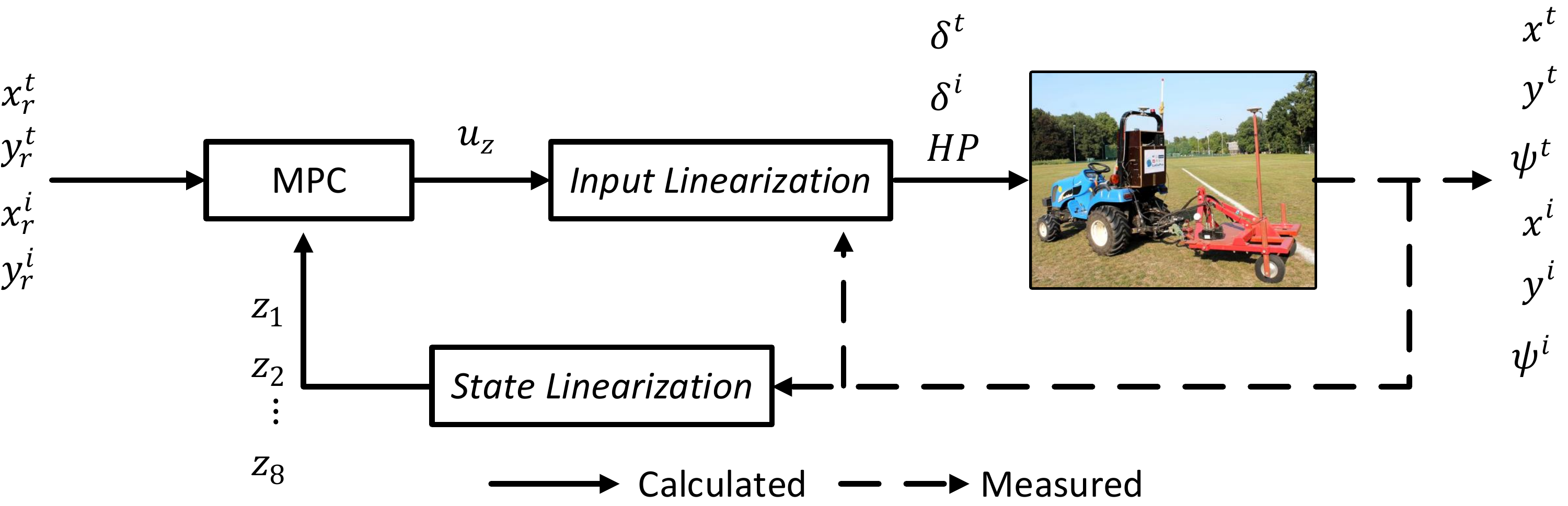}\\
  \caption{Control scheme for the LMPC by using the input-state linearization}
  \label{controlschemeIST}
\end{figure}

As can be seen from \eqref{lmpc}, the LMPC formulation is a convex optimization problem while the formulation for NMPC in \eqref{nmpc} is the constrained nonlinear optimization problem that is non-convex. Therefore, it should be noted that the required computational time for LMPC is significantly less than the one for NMPC.  The LMPC was implemented by using the Model Predictive Control toolbox in LabVIEW, which is a traditional method.

The block diagram of the control scheme for the IST-LMPC framework is shown in Fig. \ref{controlschemeIST}. As can be seen in Fig. \ref{controlschemeIST}, the generated inputs for the linear system are fed to the input linearization transformation to find proper inputs for the real-time system. Similarly, the outputs of the real-time system are fed to the state linearization transformation to calculate the states of the linear system.


\section{Experimental Results}\label{experimentalresults}

For an autonomous ground vehicle application, there are two types of reference definitions: one is a time-based trajectory and the second is a space-based trajectory. Whereas the longitudinal speed of the ground vehicle is constant in the latter, it is controlled in the former \cite{Kayacan201578,erkan2016mech}. The space-based trajectory approach is convenient in case of one vehicle in agricultural operations. If several vehicles are operating cooperatively, some of them need to be in a specific position in a specific time instant. For example, if a combine harvester and multiple tractor-trailer combinations are operating together, the tractor-trailer systems have to align with and follow the combine harvester which may vary its speed to maximally use its capacity. Therefore, the tractor-trailer systems should change their speed to get in line with and keep track of the combine harvester. This cannot be obtained with a space-based trajectory but requires the tracking of a time-based trajectory approach. Another example is the tracking of path with variable speed to adapt the machine to variable crop density. Therefore, a time-based trajectory consisting of an 8-shaped trajectory has been used as a reference signal. The 8-shaped trajectory consists of two smooth curvilinear lines and two straight lines. 

\begin{figure*}[t!]
\centering
\subfigure[]{
\includegraphics[width=2.25in]{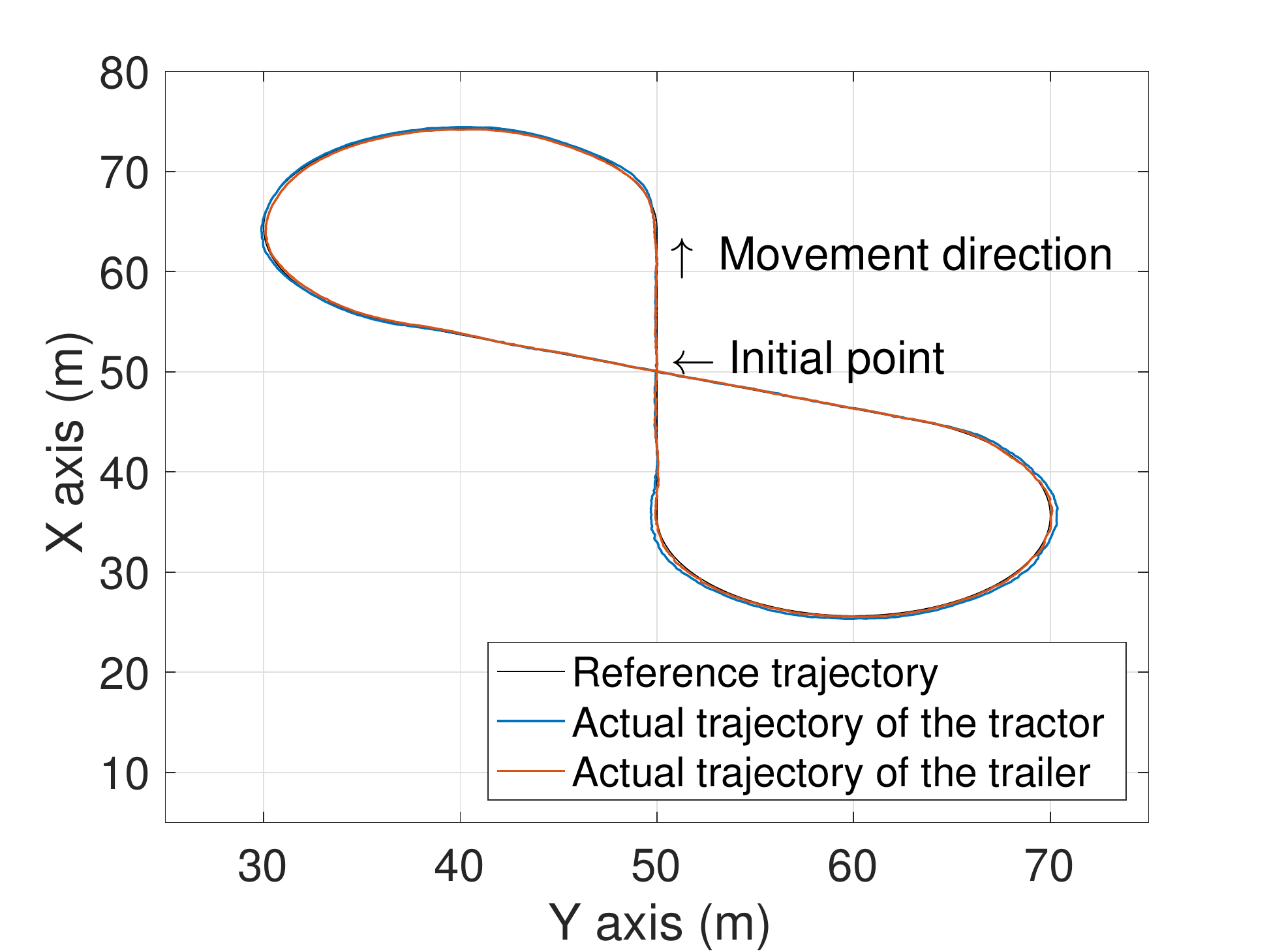}
\label{traj_NMPC}
}
\subfigure[ ]{
\includegraphics[width=2.25in]{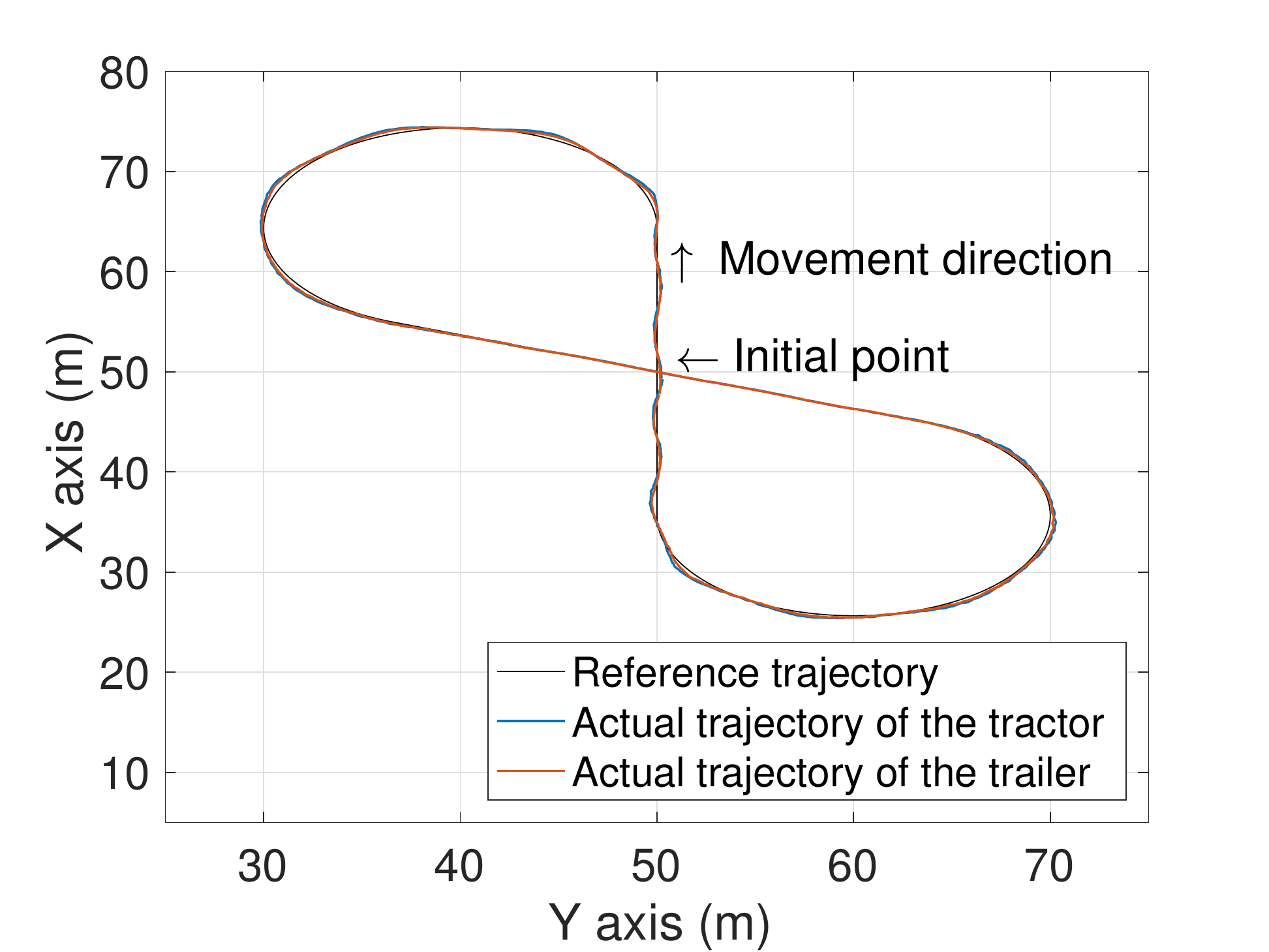}
\label{traj_LMPC}
}
\subfigure[ ]{
\includegraphics[width=2.25in]{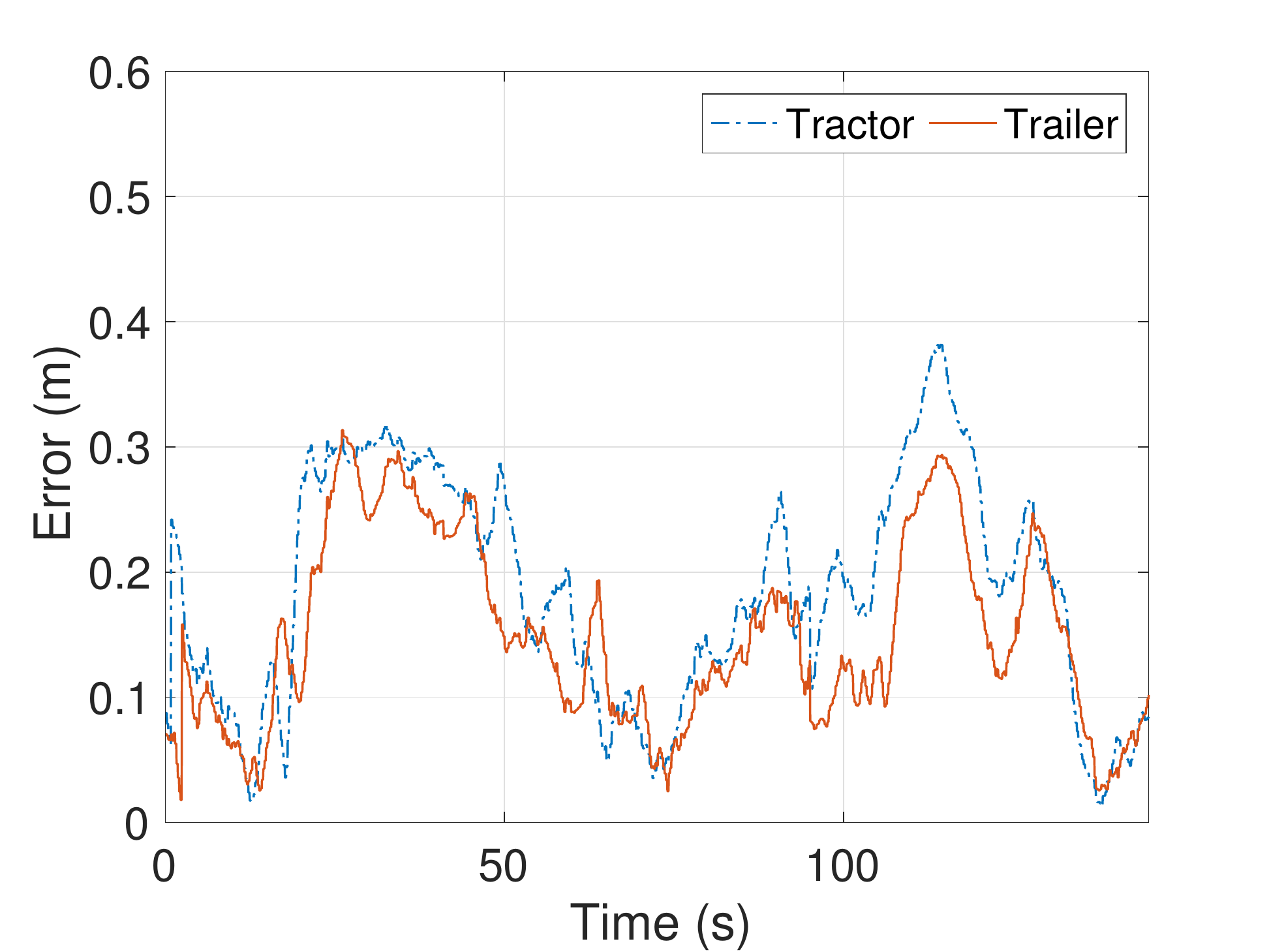}
\label{error_NMPC}
}
\subfigure[ ]{
\includegraphics[width=2.25in]{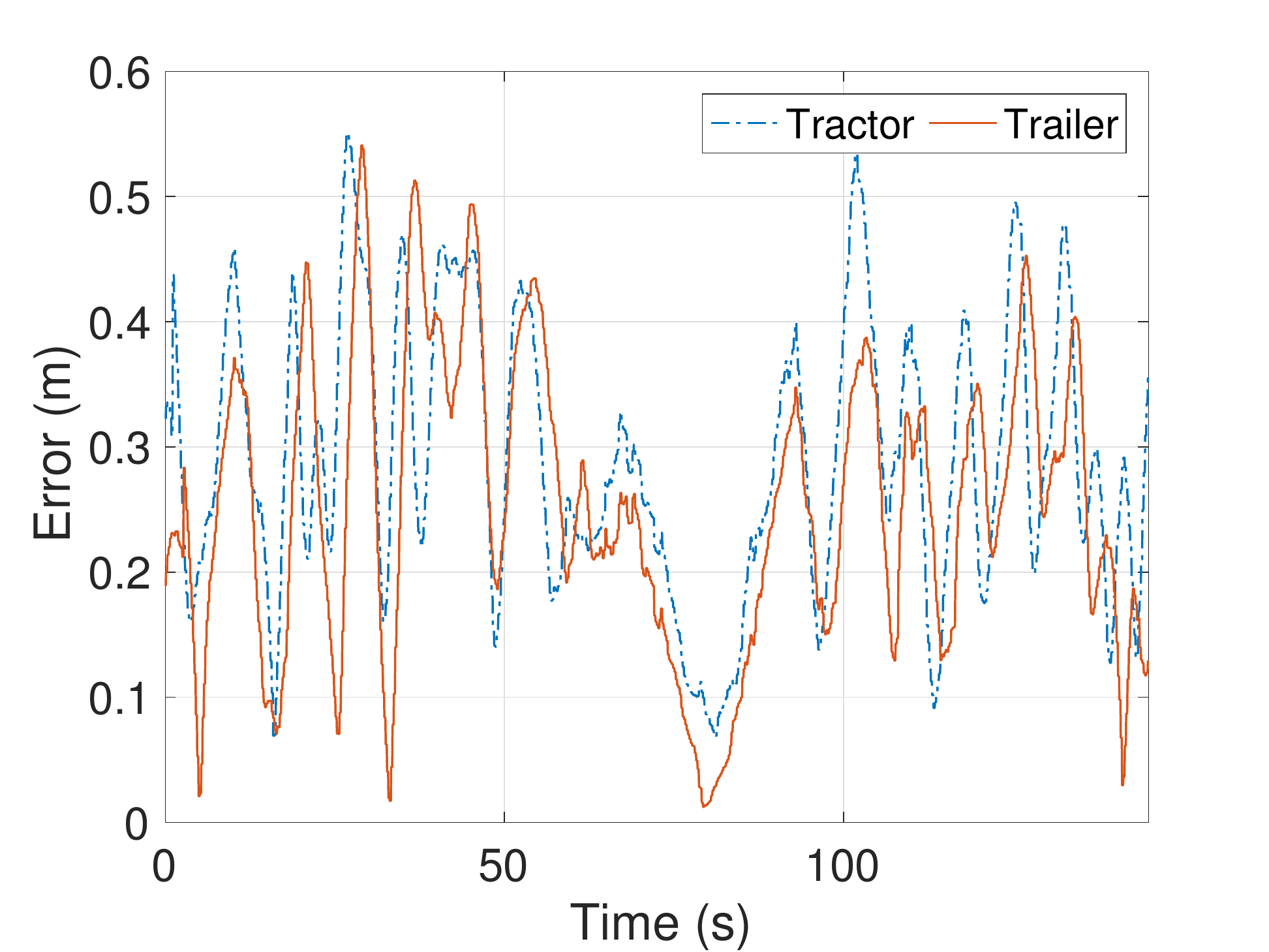}
\label{error_LMPC}
}
\subfigure[ ]{
\includegraphics[width=2.25in]{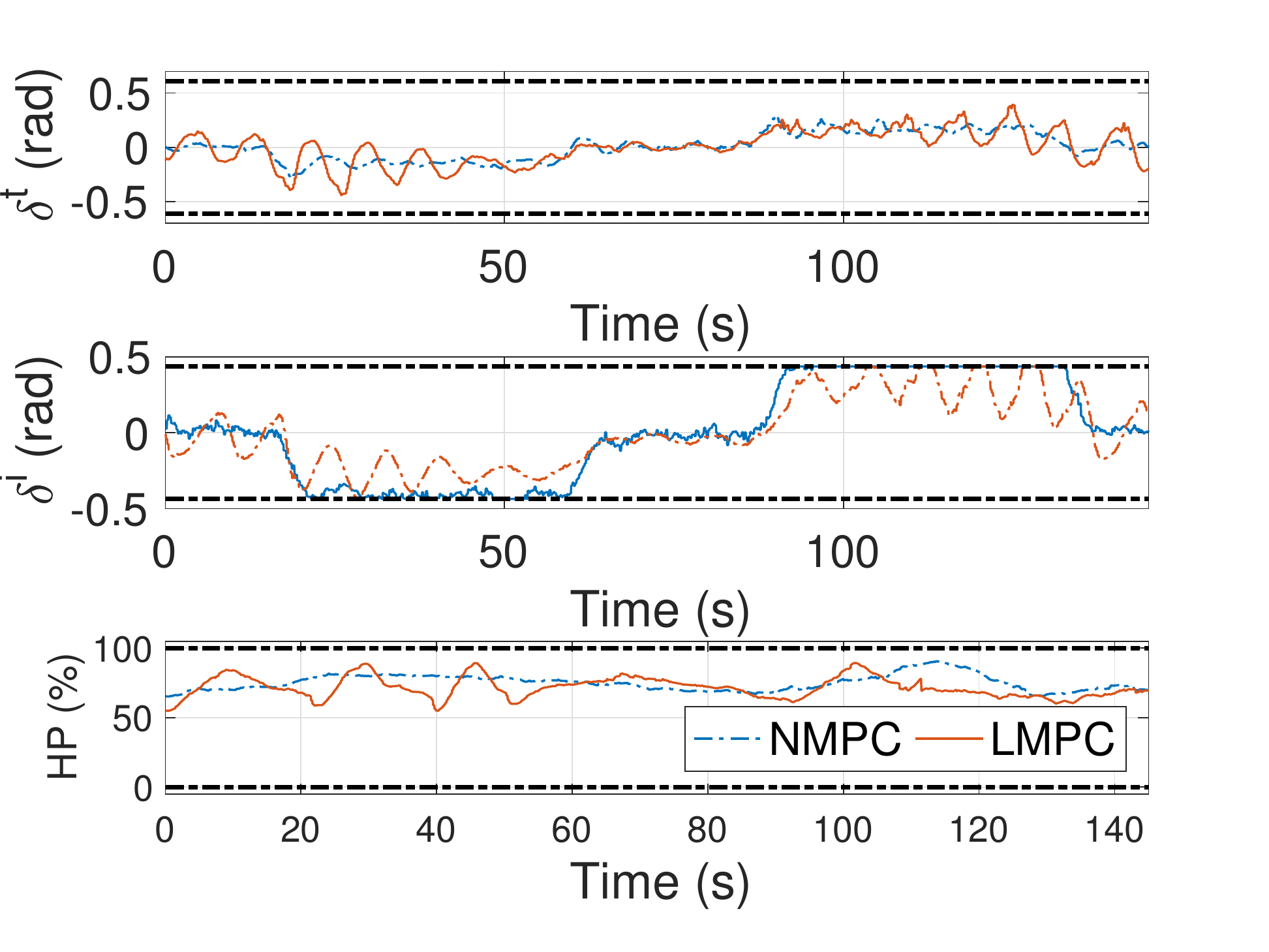}
\label{controlsignals}
}
\subfigure[ ]{
\includegraphics[width=2.25in]{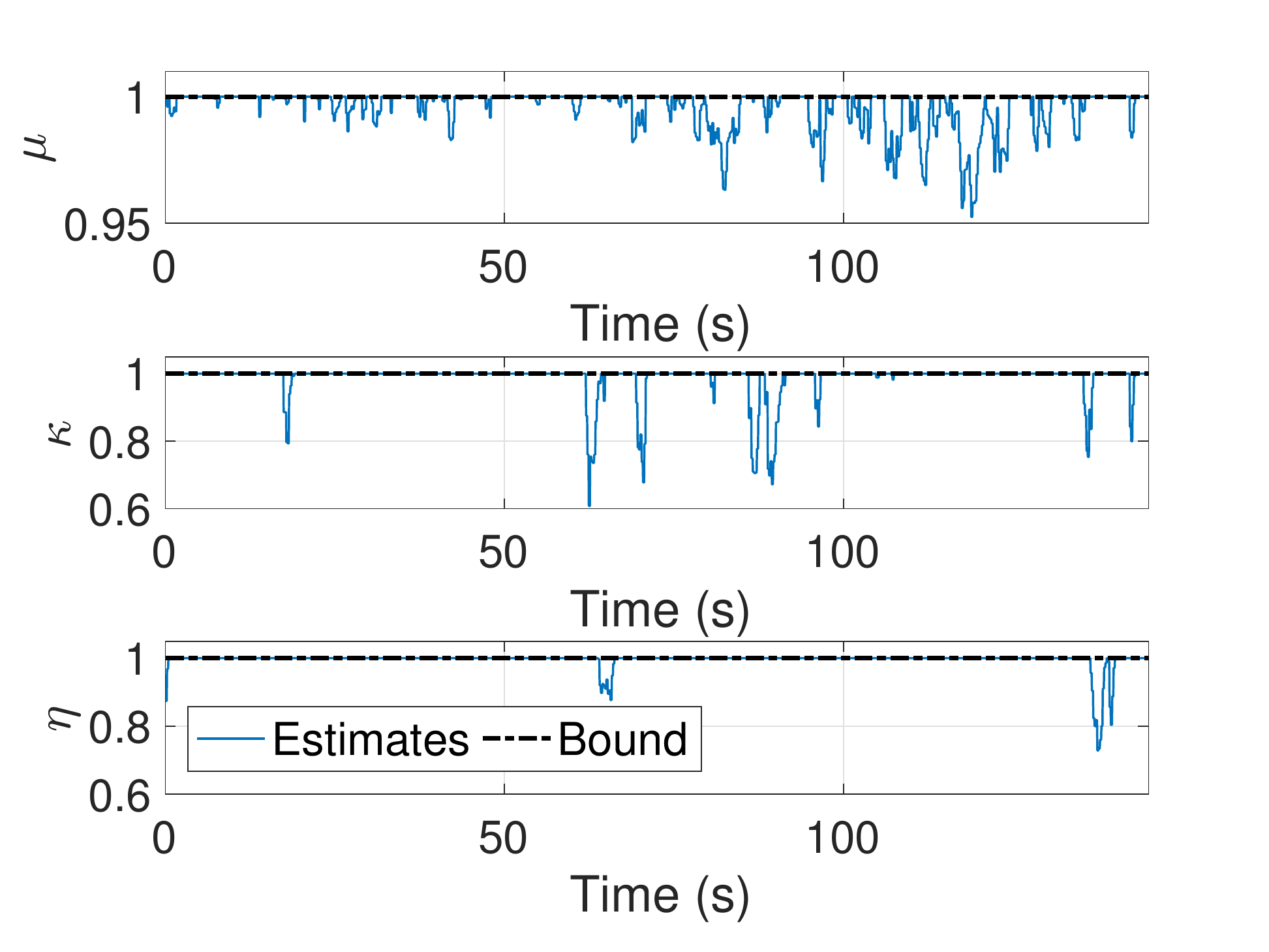}
\label{slips}
}
\caption[Optional caption for list of figures]{ Experimental results (a) Reference and actual trajectories for NMPC (b) Reference and actual trajectories for LMPC (c) Euclidean error to the reference trajectory for NMPC (d) Euclidean error to the reference trajectory for LMPC (e) Control signals (f) Traction parameters }
\label{fig_1}
\end{figure*}

Throughout the experiments, the articulated unmanned ground vehicle has faced with uneven terrain and the sampling time of the frameworks is $0.2$ second in real-time. The autonomous tractor-trailer system has succeeded in staying on-track for the NMHE-NMPC and ISL-LMPC frameworks as shown respectively in Figs. \ref{traj_NMPC} and \ref{traj_LMPC}. 

The Euclidean errors for the tractor and trailer are respectively shown for the nonlinear and linear controllers in Figs. \ref{error_NMPC} and \ref{error_LMPC}. By using the NMHE-NMPC framework, the mean values of the Euclidean errors of the tractor and trailer are obtained respectively $16.65$ cm and $10.32$ cm for the straight lines while $33.09$ cm and $25.01$ cm for the curvilinear lines. it is pointed out that that the trajectory tracking error for straight lines has been less than the one for the curvilinear lines as shown in Fig. \ref{error_NMPC}. The same framework was implemented for the space-based trajectory method in \cite{erkanCeNMPC} while the time-based one has been used in this paper. The trajectory error to the space-based trajectory was less than the one to the time-based trajectory for straight lines, while it was more than the one to the time-based trajectory for curvilinear lines. Therefore, it can be concluded that the preferred approach depends on the shape of the trajectory. Moreover, an NMHE-NMPC framework for the time-based approach was designed for agricultural vehicles in \cite{TomKraus}. It was reported that the Euclidean error was around $1$ m. This shows the superiority of our frameworks.

By using the ISL-LMPC framework, the mean values of the Euclidean errors of the tractor and the trailer have been obtained respectively $19.26$ cm and $15.27$ cm for the straight lines while $37.01$ cm and $33.33$ cm for the curvilinear lines as shown in Fig. \ref{error_LMPC}. As reported in \cite{snider2009automatic}, the linear control techniques are invalid for curvilinear trajectories. However, thanks to the ISL transformation, the ISL-LMPC framework is capable of staying on-track. When these two frameworks have been compared, it is seen that the ISL-LMPC framework has performed worse than the NMHE-NMPC framework for both the tracking of the straight and the curvilinear lines. In the ISL-LMPC framework, the traction parameters are excluded from the model and the linearization of the system is executed at every time-step. The aforementioned factors are the reasons for the degraded performance.

The outputs of the controllers, which are the steering angles references for the tractor and trailer ($\delta^{t}$, $\delta^i$), and the hydrostat position (HP) reference,  are illustrated in Fig. \ref{controlsignals}. As seen in these figures, the control signals stay within the bounds and the control signals generated by NMPC are more smooth than the ones generated by LMPC. The reason for this difference is the high non-linearity of the input transformation for the ISL-LMPC framework. Moreover, the estimated traction parameters by the NMHE are shown in Fig. \ref{slips}. The estimates stay within the bounds. 

The execution times for NMHE, NMPC and LMPC are summarized in Table I. Preparation time denotes required computation time to evaluate objective, constraints and condensing procedure till all measurements are received, while feedback time is the required computation time to compute linear term and constraints bounds in condensed QP, and to send generated signals to actuators. As seen from this table, the average computation times for NMHE and NMPC were respectively equal to $6.8575$ ms and $5.3904$ ms. Thus, the overall computation time for the NMHE-NMPC framework was equal to $12.2479$ ms. While the maximal computation time for the NMPC has been still reasonable for in real-time, the mean value of the computation time for LMPC has been $8$ times lower with $1.2330$ ms. Moreover, it is required to monitor the Karush-Kuhn-Tucker tolerances to check the optimality of the optimization problems for the NMHE and NMPC. The mean values of the KKT tolerances were respectively $7.8641$ $10^{-4}$ and $4.264$ $10^{-3}$. They are low enough to claim the optimality.
\begin{table}[h!]
\centering
\caption{Execution times of the NMHE, NMPC, and LMPC.}\label{performance}
\begin{tabular}{lccc}
  \hline
   &  NMHE   &  NMPC   &  LMPC \\
       \hline
  Preparation step & 6.0637 & 5.1745 &  1.2000 \\
  Feedback  step  & 0.7938 & 0.2159  &  0.0330  \\
  Overall     & 6.8575 & 5.3904  &  1.2330  \\
  \hline
\end{tabular}
\end{table}


\section{Conclusions}\label{conclusions}

The NMHE-NMPC and ISL-LMPC frameworks have been developed for the time-based trajectory tracking problem of an articulated unmanned ground vehicle and implemented on a real-time system. The experimental results have shown that both frameworks are capable of keeping the system on-track. Thanks to the 1-step Gauss-Newton iteration principle, the computationally efficient NMHE-NMPC framework requires a computation time of around $12$ ms, while the computation time for the ISL-LMPC framework is less than $2$ ms. This reduction on computational burden came at the price of a worse tracking error; however, the ISL-LMPC framework can be used in case of limited computation power in real-time.

Recent developments in microprocessors technology and fast solution tools for NMPC have changed the well-known paradigm in a way that the belief of using NMPC for only relatively slow dynamic systems is no longer true. The comparative results presented in this paper also show that NMPC implementations for fast robotic systems do not require enormous computation power anymore. 

\bibliography{ist-lmpc}

\begin{thebibliography}{10}
\providecommand{\url}[1]{#1}
\csname url@samestyle\endcsname
\providecommand{\newblock}{\relax}
\providecommand{\bibinfo}[2]{#2}
\providecommand{\BIBentrySTDinterwordspacing}{\spaceskip=0pt\relax}
\providecommand{\BIBentryALTinterwordstretchfactor}{4}
\providecommand{\BIBentryALTinterwordspacing}{\spaceskip=\fontdimen2\font plus
\BIBentryALTinterwordstretchfactor\fontdimen3\font minus
  \fontdimen4\font\relax}
\providecommand{\BIBforeignlanguage}[2]{{%
\expandafter\ifx\csname l@#1\endcsname\relax
\typeout{** WARNING: IEEEtran.bst: No hyphenation pattern has been}%
\typeout{** loaded for the language `#1'. Using the pattern for}%
\typeout{** the default language instead.}%
\else
\language=\csname l@#1\endcsname
\fi
#2}}
\providecommand{\BIBdecl}{\relax}
\BIBdecl

\bibitem{fao}
FAO, ``How to feed the world in 2050,'' Food and Agriculture Organization of
  the United Nations (FAO), Tech. Rep., October 2009.

\bibitem{julian1971design}
A.~Julian, ``Design and performance of a steering control system for
  agricultural tractors,'' \emph{Journal of Agricultural Engineering Research},
  vol.~16, no.~3, pp. 324--336, 1971.

\bibitem{reid1987vision}
J.~Reid and S.~Searcy, ``Vision-based guidance of an agriculture tractor,''
  \emph{Control Systems Magazine, IEEE}, vol.~7, no.~2, pp. 39--43, 1987.

\bibitem{hiremath2014laser}
S.~A. Hiremath, G.~W. Van Der~Heijden, F.~K. Van~Evert, A.~Stein, and C.~J.
  Ter~Braak, ``Laser range finder model for autonomous navigation of a robot in
  a maize field using a particle filter,'' \emph{Computers and Electronics in
  Agriculture}, vol. 100, pp. 41--50, 2014.

\bibitem{bradford1996global}
P.~W. Bradford, J.~Spilker, and P.~Enge, ``Global positioning system: theory
  and applications,'' \emph{AIAA Washington DC}, vol. 109, 1996.

\bibitem{bell2000automatic}
T.~Bell, ``Automatic tractor guidance using carrier-phase differential {GPS},''
  \emph{Computers and electronics in agriculture}, vol.~25, no. 1-2, pp.
  53--66, 2000.

\bibitem{Bevly2007}
D.~Bevly and B.~Parkinson, ``Cascaded kalman filters for accurate estimation of
  multiple biases, dead-reckoning navigation, and full state feedback control
  of ground vehicles,'' \emph{IEEE Transactions on Control Systems Technology},
  vol.~15, no.~2, pp. 199--208, March 2007.

\bibitem{Khalaji2014}
A.~Khalaji and S.~Moosavian, ``Robust adaptive controller for a tractor-trailer
  mobile robot,'' \emph{Mechatronics, IEEE/ASME Transactions on}, vol.~19,
  no.~3, pp. 943--953, June 2014.

\bibitem{Michalek2015}
M.~Michalek and M.~Kielczewski, ``The concept of passive control assistance for
  docking maneuvers with n-trailer vehicles,'' \emph{Mechatronics, IEEE/ASME
  Transactions on}, vol.~20, no.~5, pp. 2075--2084, Oct 2015.

\bibitem{Derrick}
J.~B. Derrick and D.~M. Bevly, ``Adaptive steering control of a farm tractor
  with varying yaw rate properties,'' \emph{Journal of Field Robotics},
  vol.~26, no. 6-7, pp. 519--536, 2009.

\bibitem{karkeejournal}
M.~Karkee and B.~L. Steward, ``Study of the open and closed loop
  characteristics of a tractor and a single axle towed implement system,''
  \emph{Journal of Terramechanics}, vol.~47, no.~6, pp. 379 -- 393, 2010.

\bibitem{7505627}
H.~Li and W.~Yan, ``Model predictive stabilization of constrained underactuated
  autonomous underwater vehicles with guaranteed feasibility and stability,''
  \emph{IEEE/ASME Transactions on Mechatronics}, vol.~22, no.~3, pp.
  1185--1194, June 2017.

\bibitem{Kayacan2018}
E.~Kayacan, E.~Kayacan, I.-M. Chen, H.~Ramon, and W.~Saeys, \emph{On the
  Comparison of Model-Based and Model-Free Controllers in Guidance, Navigation
  and Control of Agricultural Vehicles}.\hskip 1em plus 0.5em minus 0.4em\relax
  Cham: Springer International Publishing, 2018, pp. 49--73.

\bibitem{Backman2012}
J.~Backman, T.~Oksanen, and A.~Visala, ``Navigation system for agricultural
  machines: Nonlinear model predictive path tracking,'' \emph{Computers and
  Electronics in Agriculture}, vol.~82, pp. 32 -- 43, 2012.

\bibitem{TomKraus}
T.~Kraus, H.~Ferreau, E.~Kayacan, H.~Ramon, J.~D. Baerdemaeker, M.~Diehl, and
  W.~Saeys, ``Moving horizon estimation and nonlinear model predictive control
  for autonomous agricultural vehicles,'' \emph{Computers and Electronics in
  Agriculture}, vol.~98, pp. 25 -- 33, 2013.

\bibitem{erkanCeNMPC}
E.~Kayacan, E.~Kayacan, H.~Ramon, and W.~Saeys, ``Learning in centralized
  nonlinear model predictive control: Application to an autonomous
  tractor-trailer system,'' \emph{Control Systems Technology, IEEE Transactions
  on}, vol.~23, no.~1, pp. 197--205, Jan 2015.

\bibitem{erkanDeNMPC}
------, ``Robust tube-based decentralized nonlinear model predictive control of
  an autonomous tractor-trailer system,'' \emph{Mechatronics, IEEE/ASME
  Transactions on}, vol.~20, no.~1, pp. 447--456, Feb 2015.

\bibitem{erkanDiNMPC}
------, ``Distributed nonlinear model predictive control of an autonomous
  tractor–trailer system,'' \emph{Mechatronics}, vol.~24, no.~8, pp. 926 --
  933, 2014.

\bibitem{karkeephd}
M.~Karkee, ``Modeling, identification and analysis of tractor and single axle
  towed implement system,'' Ph.D. dissertation, Iowa State University, 2009.

\bibitem{erkan2016mech}
E.~Kayacan, H.~Ramon, and W.~Saeys, ``Robust trajectory tracking error
  model-based predictive control for unmanned ground vehicles,''
  \emph{IEEE/ASME Transactions on Mechatronics}, vol.~21, no.~2, pp. 806--814,
  April 2016.

\bibitem{erkanmodelleme}
E.~Kayacan, E.~Kayacan, H.~Ramon, and W.~Saeys, ``Nonlinear modeling and
  identification of an autonomous tractor–trailer system,'' \emph{Computers
  and Electronics in Agriculture}, vol. 106, no.~0, pp. 1 -- 10, 2014.

\bibitem{Rao2003}
C.~Rao, J.~Rawlings, and D.~Mayne, ``Constrained state estimation for nonlinear
  discrete-time systems: stability and moving horizon approximations,''
  \emph{Automatic Control, IEEE Transactions on}, vol.~48, no.~2, pp. 246--258,
  2003.

\bibitem{Ferreau}
H.~Ferreau, T.~Kraus, M.~Vukov, W.~Saeys, and M.~Diehl, ``High-speed moving
  horizon estimation based on automatic code generation,'' in \emph{Decision
  and Control (CDC), 2012 IEEE 51st Annual Conference on}, 2012, pp. 687--692.

\bibitem{Mayne}
D.~Mayne, J.~Rawlings, C.~Rao, and P.~Scokaert, ``Constrained model predictive
  control: Stability and optimality,'' \emph{Automatica}, vol.~36, no.~6, pp.
  789 -- 814, 2000.

\bibitem{Falcone2007}
P.~Falcone, F.~Borrelli, J.~Asgari, H.~Tseng, and D.~Hrovat, ``Predictive
  active steering control for autonomous vehicle systems,'' \emph{Control
  Systems Technology, IEEE Transactions on}, vol.~15, no.~3, pp. 566--580, May
  2007.

\bibitem{Diehl2005}
M.~Diehl, H.~G. Bock, and J.~P. Schlöder, ``A real-time iteration scheme for
  nonlinear optimization in optimal feedback control,'' \emph{SIAM Journal on
  Control and Optimization}, vol.~43, no.~5, pp. 1714--1736, 2005.

\bibitem{Houska2011a}
B.~Houska, H.~J. Ferreau, and M.~Diehl, ``Acado toolkit—an open-source
  framework for automatic control and dynamic optimization,'' \emph{Optimal
  Control Applications and Methods}, vol.~32, no.~3, pp. 298 -- 312, 2011.

\bibitem{Diehl}
M.~Diehl, H.~Bock, J.~P. Schlöder, R.~Findeisen, Z.~Nagy, and F.~Allgower,
  ``Real-time optimization and nonlinear model predictive control of processes
  governed by differential-algebraic equations,'' \emph{Journal of Process
  Control}, vol.~12, no.~4, pp. 577 -- 585, 2002.

\bibitem{Ferreau2014}
H.~J. Ferreau, C.~Kirches, A.~Potschka, H.~G. Bock, and M.~Diehl, ``qpoases: a
  parametric active-set algorithm for quadratic programming,''
  \emph{Mathematical Programming Computation}, vol.~6, no.~4, pp. 327--363, Dec
  2014.

\bibitem{slotine}
J.-J.~E. Slotine, W.~Li \emph{et~al.}, \emph{Applied nonlinear control}.\hskip
  1em plus 0.5em minus 0.4em\relax Prentice-Hall Englewood Cliffs, NJ, 1991,
  vol. 199, no.~1.

\bibitem{Kayacan201578}
E.~Kayacan, E.~Kayacan, H.~Ramon, and W.~Saeys, ``Towards agrobots:
  Identification of the yaw dynamics and trajectory tracking of an autonomous
  tractor,'' \emph{Computers and Electronics in Agriculture}, vol. 115, pp. 78
  -- 87, 2015.

\bibitem{snider2009automatic}
J.~M. Snider, ``Automatic steering methods for autonomous automobile path
  tracking,'' \emph{Robotics Institute, Pittsburgh, PA, Tech. Rep.
  CMU-RITR-09-08}, 2009.

\end{thebibliography}
\bibliographystyle{IEEEtran}

\begin{IEEEbiography}[{\includegraphics[width=1in,height=1.25in,clip,keepaspectratio]{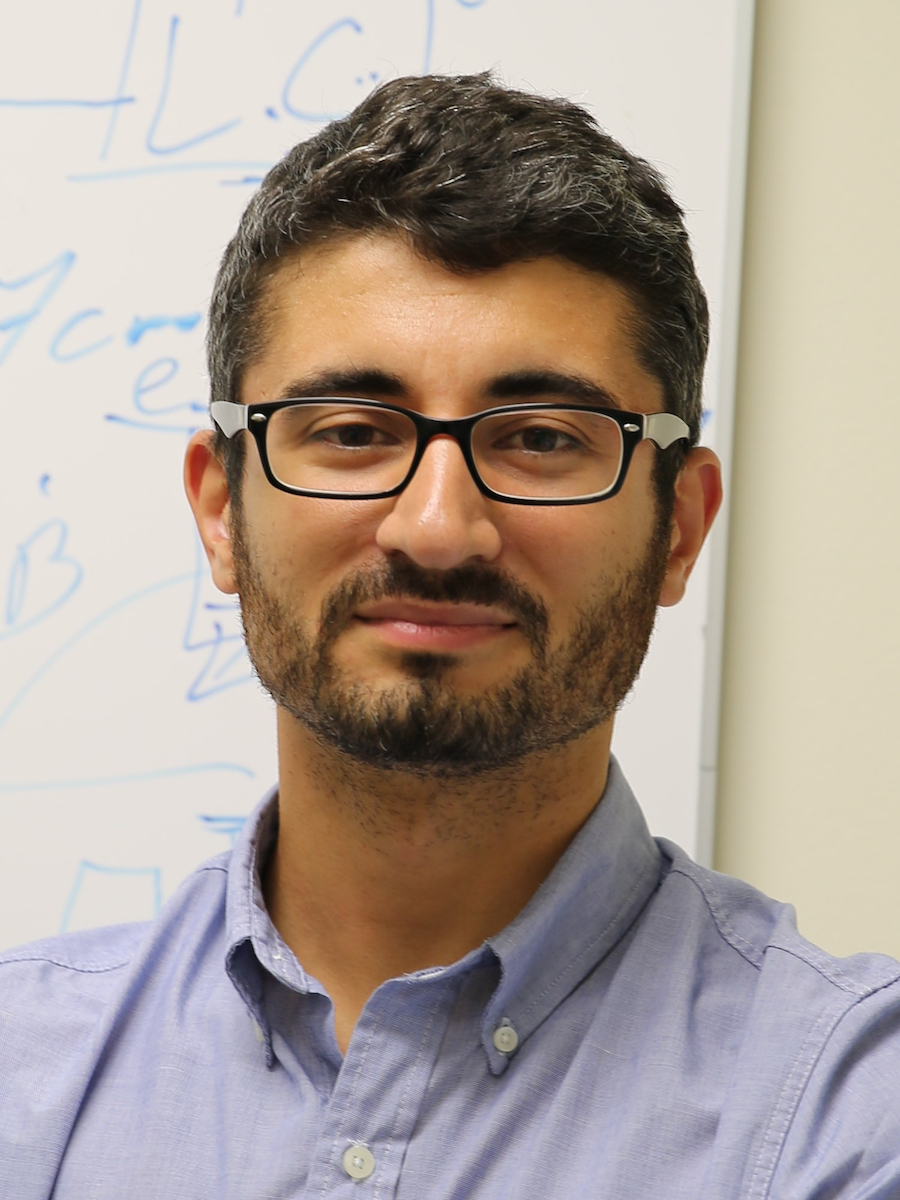}}]{Erkan Kayacan} (S\textquoteright 12 M\textquoteright 16) was born in Istanbul, Turkey, on April 17, 1985. He received the B.Sc. degree in mechanical engineering and the M.Sc. degree in system dynamics and control from Istanbul Technical University, Istanbul, Turkey in 2008 and 2010, respectively. He received the Ph.D. degree in Mechatronics, Biostatistics and Sensors from University of Leuven (KU Leuven), Leuven, Belgium in 2014.

He is currently a Postdoctoral Associate with Senseable City Laboratory and Computer Science and Artificial Intelligence Laboratory, Massachusetts Institute of Technology, MA, USA. Prior to MIT, he was a Postdoctoral Researcher with Delft Center for Systems and Control (DCSC), Delft University of Technology, The Netherlands and Distributed Autonomous Systems lab, University of Illinois at Urbana-Champaign, IL, USA. His research interests include  real-time optimization-based control and estimation methods, nonlinear control theory, learning algorithms and machine learning with a heavy emphasis on applications to autonomous systems and field robotics.

Dr. Kayacan is a recipient of the Best Systems Paper Award at Robotics: Science and Systems (RSS) in 2018.
\end{IEEEbiography}

\begin{IEEEbiography}[{\includegraphics[width=1in,height=1.25in,clip,keepaspectratio]{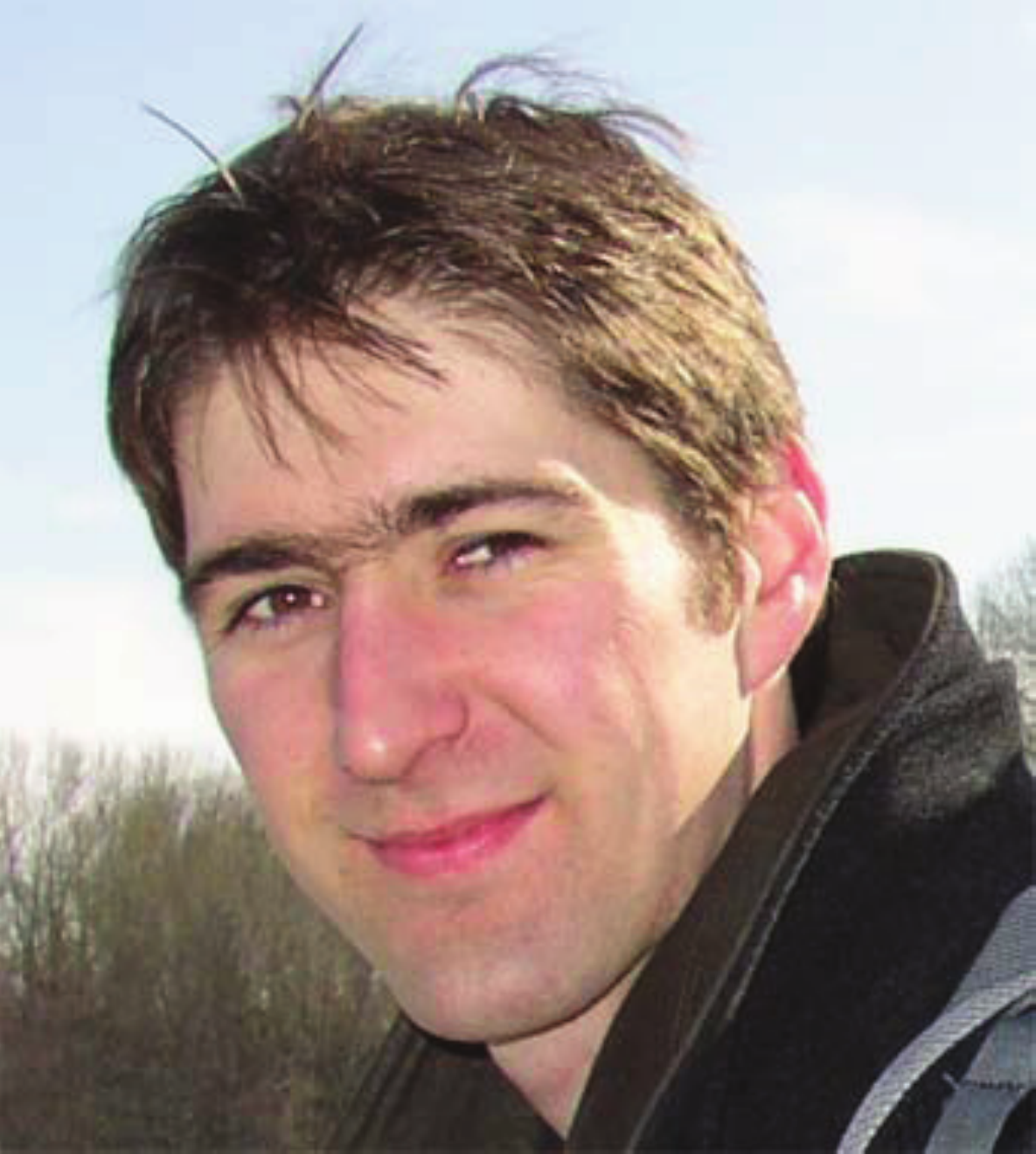}}]{Wouter Saeys} received the M.Sc degree in Bioscience Engineering from University of Leuven (KU Leuven), Leuven, Belgium in 2002. On the basis of his Master’s thesis, he was awarded  the engineering prize by the Royal Flemish Society of Engineers (KVIV). In 2006, he received the Ph.D. in Bioscience Engineering from KU Leuven, Leuven, Belgium.

Since 2017, he is an Associate Professor at the Biosystems Department of KU Leuven, where he leads a group focusing on Agrofood Mechatronics. His main research interests include agricultural automation and robotics, chemometrics, light transport modelling and optical characterization of biological materials. He has supervised 20 PhDs and is (co-)author of over 145 peer reviewed journal articles.
\end{IEEEbiography}

\begin{IEEEbiography}[{\includegraphics[width=1in,height=1.25in,clip,keepaspectratio]{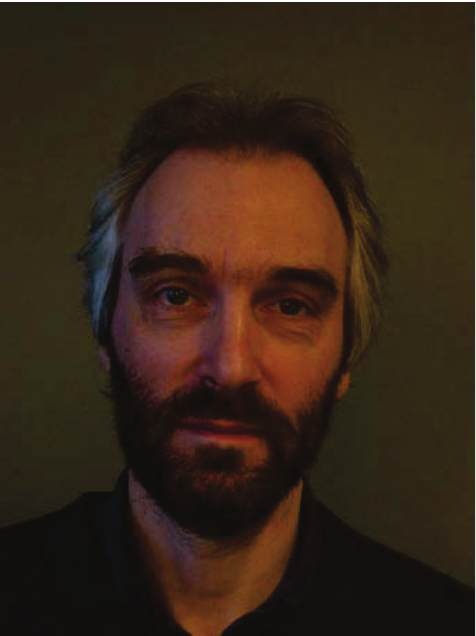}}]{Herman Ramon} received the M.Sc. degree in bioscience engineering from Gent University, Gent, Belgium and the Ph.D. degree in biological sciences from the University of Leuven (KU Leuven), Leuven, Belgium, in 1993.

He is currently a Professor with the Faculty of Bioscience Engineering, KU Leuven, lecturing on field robotics, system dynamics, applied mechanics and mathematical biology. His current research interests include mechatronic design on agricultural machinery, theoretical and experimental study of particulate systems from fundamental research in single-cell and tissue dynamics to granular mechanics and liquid-like flow in agricultural and industrial processes. He has (co-)authored more than 200 peer reviewed journal articles.
 \end{IEEEbiography}

\begin{IEEEbiography}[{\includegraphics[width=1in,height=1.25in,clip,keepaspectratio]{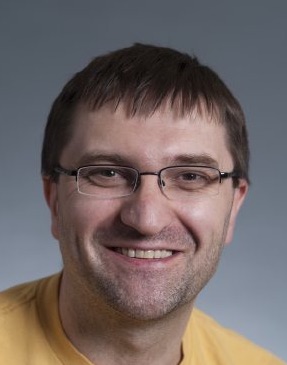}}]{Calin Belta} (M\textquoteright 03-SM\textquoteright 11-F\textquoteright 17) received the B.Sc. and M.Sc. degrees in control and computer science from the Technical University of Iasi, Iasi, Romania, in 1995 and 1996, respectively, the M.Sc. degree in electrical engineering from Louisiana State University, Baton Rouge, LA, USA, in 1999, and the M.Sc. and Ph.D. degrees in mechanical engineering from the University of Pennsylvania, Philadelphia, PA, USA, in 2001 and 2003, respectively.

He is currently Professor with the Department of Mechanical Engineering, Boston University (BU), Boston, MA, USA, where he holds the Tegan family Distinguished Faculty Fellowship. He is the Director of the BU Robotics Laboratory and of the Center for Autonomous and Robotic Systems. He is also affiliated with the Department of Electrical and Computer Engineering, the Division of Systems Engineering, the Center for Information and Systems Engineering, and the Bioinformatics Program. His research interests include dynamics and control theory, with particular emphasis on hybrid and cyber-physical systems, formal synthesis and verification, and applications in robotics and systems biology.

Prof. Belta received the Air Force Office of Scientific Research Young Investigator Award and the National Science Foundation CAREER Award.
\end{IEEEbiography}

\begin{IEEEbiography}[{\includegraphics[width=1in,height=1.25in,clip,keepaspectratio]{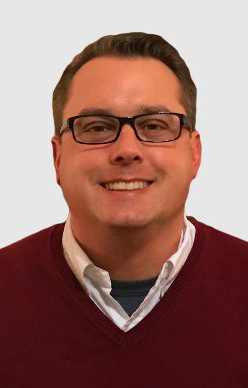}}]{Joshua M. Peschel}(IEEE Member) is an Assistant Professor of Agricultural and Biosystems Engineering, Black \& Veatch Faculty Fellow, and Director of the Human-Infrastructure Interaction Lab at Iowa State University. He received the B.S. in Biological Systems Engineering, M.S. in Biological and Agricultural Engineering, and Ph.D. in Computer Science in 2001, 2004, and 2012, all from Texas A\&M University. His research interests focus on innovative sensing and sense-making for smart agricultural, natural, and urban systems.
\end{IEEEbiography}

\clearpage
\end{document}